\documentclass[aps,prd,preprint,singlecolumn,superscriptaddress,nofootinbib,longbibliography]{revtex4-2}
\usepackage[T1]{fontenc}
\usepackage[utf8]{inputenc}
\usepackage{lmodern}
\usepackage{microtype}
\usepackage{amsmath,amssymb,bm}
\usepackage{mathtools}
\usepackage{xcolor}
\usepackage{booktabs}
\usepackage{enumitem}
\usepackage{hyperref}
\usepackage{tikz}
\usepackage{pgfplots}
\usepgfplotslibrary{fillbetween,groupplots}
\pgfplotsset{compat=1.18}

\definecolor{navy}{RGB}{23,54,93}
\newcommand{\rev}[1]{#1}
\newcommand{\mq}{\mathfrak q}
\newcommand{\Ain}{A_{\mathrm{in}}}
\newcommand{\Bin}{B_{\mathrm{in}}}
\newcommand{\Pu}{\mathcal P_u}
\newcommand{\Pv}{\mathcal P_v}
\newcommand{\dd}{\mathrm d}
\newcommand{\ee}{\mathrm e}
\newcommand{\cD}{\mathcal D}
\newcommand{\cJ}{\mathcal J}

\hypersetup{colorlinks=true,linkcolor=blue,citecolor=blue,urlcolor=blue}
\setlist[itemize]{leftmargin=2em}
\preprint{CYCU-HEP-26-08}

\begin{document}
\title{Polyakov backreaction in large $D$ discretely self-similar collapse}
\author{Sheng-Hong Lai}
\email{shlai@cycu.edu.tw}
\affiliation{Department of Physics and Center for High Energy Physics, Chung Yuan Christian University, Chung Li City, Taiwan}
\author{Wen-Yu Wen}
\email{wenw@cycu.edu.tw}
\affiliation{Department of Physics and Center for High Energy Physics, Chung Yuan Christian University, Chung Li City, Taiwan}

\begin{abstract}
We study anomaly-induced semiclassical backreaction on an analytic family of discretely self-similar solutions of Einstein--Klein--Gordon theory at large spacetime dimension. Horizon regularity fixes a null-coordinate Floquet exponent and selects a unique regular quantum state. The Polyakov auxiliary field is affine-periodic rather than periodic, with a monodromy that is determined independently by a global zero-mode condition and by the conformal-coordinate scaling. Spherical reduction introduces a quantum coupling that grows toward the center, so the periodic first-order expansion is nonuniform. We construct the resulting nonlinear inner problem and find a critical dilaton surface. Generic crossings of this surface are singular, whereas matching to the regular zero-mass exterior selects a narrow, high-curvature throat. The anomaly also separates the effective null-energy boundaries from the classical matter flux and zero-curvature curves. These results show that the minimally coupled Polyakov model does not yield a generic smooth, exactly self-similar completion of the classical center.

\end{abstract}
\maketitle

\section{Introduction}

Critical gravitational collapse occupies a distinctive place in classical general relativity. In spherical symmetry, and for a massless scalar field minimally coupled to Einstein gravity, fine-tuned initial data produce a threshold solution that sits exactly between complete dispersal and black-hole formation~\cite{Choptuik:1993,GChristo:1986,GChristo:1991}. The critical solution discovered by Choptuik is discretely self-similar (DSS): after a discrete echoing period $\Delta$, the geometry and the matter field repeat themselves upon a simultaneous rescaling of time and radius. The associated spacetime contains a self-similar horizon (SSH), an ingoing null surface on which the discrete lattice vector becomes null, and a naked singularity at the future endpoint of that surface~\cite{Gundlach:1999,MartinGarcia:2007,MartinGarcia:2003}. Between the center and the SSH the lattice vector is timelike; between the SSH and the outgoing Cauchy horizon it is spacelike. The geometry therefore behaves as a conformal spacetime crystal, with a single fundamental domain that is repeated under $\tau\mapsto\tau+\Delta$.

Analytic control of this structure has remained limited. Einstein gravity in four dimensions has no small dimensionless coupling, and the critical solution has been constructed mainly by imposing DSS as a boundary-value problem and solving the resulting nonlinear system numerically~\cite{MartinGarcia:2003,Gundlach:1997}. A recent development changes that situation at large spacetime dimension $D$. Using $\epsilon=1/(D-1)$ as an expansion parameter~\cite{Emparan:2013a,Emparan:2013b,Emparan:2014,Bhattacharyya:2016}, Ecker, Ecker and Grumiller constructed an infinite analytic family of DSS solutions of the Einstein--massless-Klein--Gordon equations~\cite{Ecker:2026}. At leading order the entire solution is encoded in a single periodic function $\beta(\tau)$. Continuously self-similar (CSS) members of the family correspond to constant $\beta$, while nontrivial periodic $\beta$ generate discrete echoing. The solutions are regular throughout the past patch, from the regular center out to the SSH, and they terminate in a naked central singularity in the same qualitative manner as Choptuik's critical spacetime.

Large $D$ gravity is known to admit effective two-dimensional descriptions~\cite{Emparan:2013a,Emparan:2013b}. After spherical reduction on the unit $S^n$, with $n=D-2$, the remaining radial--temporal sector is a two-dimensional dilaton-gravity theory coupled to the $s$-wave scalar~\cite{Grumiller:2002,Cavaglia:1999}. This is a natural setting in which to ask how quantum backreaction modifies the SSH and the future central singularity. The simplest anomaly-induced model is the Polyakov action for two-dimensional conformal matter~\cite{Polyakov:1981,Polyakov:1987}. In two-dimensional black-hole physics that action is responsible for Hawking flux, for the trace anomaly, and for a number of exactly solvable semiclassical models~\cite{CGHS:1992,RST:1992,Strominger:1995,Giddings:1994,Fabbri:2005}. It is therefore a controlled first laboratory in which to test whether the classical DSS center admits a smooth, exactly self-similar quantum completion.

A naïve strategy would be to solve a periodic auxiliary-field equation on the classical background and to treat the resulting Polyakov tensor as a first-order source for a periodic metric correction. The analysis below shows that this program must be revised on several independent grounds.

\begin{enumerate}
\item The auxiliary field is affine-periodic rather than periodic. Integrating the conservative equation for $\chi$ over one echo produces a nonzero source zero-mode. Strict periodicity is obstructed and the consistent solution has a constant monodromy $C$ per echo.
\item Regularity at both the center and the SSH fixes the homogeneous Polyakov state uniquely. In regular null Floquet coordinates the SSH is the nondegenerate surface $v=0$. Smoothness there, together with even regularity at the center, eliminates nonconstant chiral state functions and selects $\chi=-2\rho+\chi_*$. 
\item The monodromy obtained from the global zero-mode condition coincides with the monodromy implied by the affine scaling of the conformal factor. The two independent calculations give the identity $C=2(1-\lambda)$, where $\lambda$ is the Floquet exponent fixed by SSH regularity.
\item The gravitational coupling of the Polyakov tensor to the spherical dilaton is $\widehat\kappa/\Phi$. Because $\Phi=r^n$ vanishes at the center, this effective coupling grows under every echo. A periodic first-order expansion at fixed quantum coupling is therefore nonuniform.
\item The resulting inner problem develops a critical kinetic determinant at a finite dilaton value $\Phi_c$. Generic crossings of that surface are curvature singularities. Matching to the regular, zero-mass exterior selects instead a narrow high-curvature throat, or cornucopion~\cite{Banks:1992,Lowe:1993,Thorlacius:1995}, before the determinant vanishes.
\end{enumerate}

These conclusions are specific to the minimally coupled intrinsic two-dimensional Polyakov model. The true $s$-wave reduction of higher-dimensional quantum matter generally produces additional dilaton-dependent anomaly terms---the dimensional-reduction anomaly~\cite{Frolov:2000,Balbinot:1999,Nojiri:1999}. However, we do not assert that the throat found below survives the inclusion of these additional dilaton-dependent anomaly terms, nor that it censors the classical singularity in the sense of cosmic censorship.

The organization is as follows. Section~2 reviews the leading-order large-$D$ DSS background and the classical coincidence between null-energy and zero-curvature curves. Section~3 constructs the spherically reduced theory, localizes the Polyakov action, derives the auxiliary monodromy, and selects the SSH-regular quantum state in conformal gauge. Section~4 writes the closed semiclassical system, identifies the critical kinetic surface, and explains why a uniform outer expansion fails. Section~5 constructs the frozen-phase inner problem, matches it to the regular exterior, and obtains the throat. Section~6 analyzes the splitting of classical null-energy curves and the negative-energy layer surrounding the throat. Section~7 discusses the limitations of the model and the calculations that come next. Technical checks and boundary-layer derivations are collected in the appendices.

\section{Classical large \texorpdfstring{$D$}{D} discretely self-similar background}

\subsection{Coordinates, fields, and the leading-order metric}

Following Ref.~\cite{Ecker:2026}, we work in the past patch of the critical spacetime and introduce the self-similar coordinates
\begin{equation}
n:=D-2,\qquad \tau=-\ln(-t),\qquad x=-\frac{r}{t},\qquad \epsilon=\frac{1}{D-1}.
\end{equation}
The range of interest is $0\le x\le1$, with $x=0$ the regular center and $x=1$ the SSH. One echo corresponds to $\tau\mapsto\tau+\Delta$. In the explicit family used below the next-to-leading large-$D$ consistency condition selects $\Delta=1$, and we adopt that normalization throughout.

The leading radial--temporal metric is
\begin{equation}
\dd s_0^2=h_{ab}\dd x^a\dd x^b=\ee^{-2\tau}\left[(x^2-f_0^2)\dd\tau^2-2x\dd\tau\dd x+\dd x^2\right],\qquad 0\le x\le1,
\end{equation}
with
\begin{equation}
\Pi_0(\tau,x)=\frac{\beta(\tau)}{\sqrt{1+\beta^2(\tau)x^2}},\quad
\Omega_0(\tau,x)=\Pi_0^2(\tau,x),\quad
f_0(\tau,x)=\sqrt{\frac{1+\beta^2(\tau)x^2}{1+\beta^2(\tau)}}.
\end{equation}
A subscript $0$ denotes a leading-order quantity. The function $f_0$ is normalized so that $f_0(\tau,1)=1$, which places the SSH at $x=1$. As a toy model, we use the explicit unit-period example of Ref.~\cite{Ecker:2026},
\begin{equation}
\beta(\tau)=\cos(2\pi\tau)+\frac{\sin(6\pi\tau)}{15.9476},\qquad \beta(\tau+1)=\beta(\tau).
\end{equation}
The numerical coefficient is chosen so that the next-to-leading consistency condition of the large-$D$ expansion is satisfied at $\Delta=1$. The same qualitative conclusions hold for any periodic $\beta$ with isolated simple zeros, vanishing mean, and the half-period antisymmetry $\beta(\tau+\Delta/2)=-\beta(\tau)$ that characterizes the family studied in Ref.~\cite{Ecker:2026}.

It is convenient to write
\begin{equation}
Q(\tau,x):=1+\beta^2(\tau)x^2,\qquad B(\tau):=1+\beta^2(\tau),\qquad f_0=\sqrt{Q/B}.
\end{equation}
For a general lapse function $f(\tau,x)$ in the ansatz (2), the intrinsic two-dimensional scalar curvature is
\begin{equation}
R[h]=-\frac{2\ee^{2\tau}}{f}\,\partial_x^2 f.
\end{equation}
On the explicit background this reduces to
\begin{equation}
R_0(\tau,x)=-\frac{2\ee^{2\tau}\beta^2(\tau)}{[1+\beta^2(\tau)x^2]^2}.
\end{equation}
The curvature is everywhere nonpositive in the past patch, vanishes on the zeros of $\beta$, and is largest in magnitude at the center.

The two-dimensional line element factorizes into radial null directions,
\begin{equation}
\dd s_0^2=\ee^{-2\tau}\bigl[\dd x-(x+f_0)\dd\tau\bigr]\bigl[\dd x-(x-f_0)\dd\tau\bigr].
\end{equation}
This factorization motivates the radial null derivatives
\begin{equation}
D_\pm:=\partial_\tau+(x\pm f)\partial_x,
\end{equation}
which become the basic transport operators of the conformal-coordinate construction in section~3.

\subsection{Matter variables and classical energy conditions}

In the normalization of Ref.~\cite{Ecker:2026} the first-order matter variables are
\begin{equation}
\Pi=\frac{(\partial_\tau+x\partial_x)\psi}{\sqrt n\,f},\qquad
\Psi=\frac{\partial_x\psi}{\sqrt n\,x},\qquad
D_\pm\psi=\sqrt n\,f(\Pi\pm x\Psi).
\end{equation}
The classical radial null energy condition (NEC) is saturated on the two matter-flux curves
\begin{equation}
\Pi-x\Psi=0,\qquad \Pi+x\Psi=0.
\end{equation}
For a minimally coupled massless scalar, the trace-reversed form of the $D$-dimensional Einstein equation implies
\begin{equation}
R_D=-n\ee^{2\tau}a^{-2}(\Pi-x\Psi)(\Pi+x\Psi),
\end{equation}
where the prefactor is $n=D-2$ and $a(t,r)$ is the radial metric function in areal gauge, $\dd s_2^2=-\alpha^2\dd t^2+a^2\dd r^2$. The two classical NEC-saturation curves therefore coincide with the $D$-dimensional zero-curvature curves.

At a simple zero $\beta(\tau_0)=0$ the two branches admit the local expansion
\begin{equation}
\tau_\pm^{\mathrm{cl}}(x)=\tau_0\pm\epsilon x-\frac{\epsilon^2x^2}{2}
-\frac{\epsilon^2x^4}{8\beta'(\tau_0)}
\left[\beta^{(4)}+10\beta'''-6(\beta')^3-55\beta'\right]_{\tau_0}+O(\epsilon^3).
\end{equation}
Here $\tau_+^{\mathrm{cl}}$ (the upper branch for $x>0$) solves $\Pi-x\Psi=0$, whereas $\tau_-^{\mathrm{cl}}$ solves $\Pi+x\Psi=0$. Their relative opening angle is $2\epsilon+O(\epsilon^3)$, while the common $O(\epsilon^2)$ term produces the first bending of the two curves away from the vertex. Both effects are parametrically small at large $D$. This will be important later: the leading quantum splitting of the same curves scales with a different combination of $\epsilon$ and the quantum coupling, and can dominate the classical $O(\epsilon^2)$ bending while the outer quantum expansion is still perturbative.

Two further classical properties will be used repeatedly. First, the leading-order fields (3) are regular for all $\tau\in\mathbb R$ and all $x\in[0,1]$. They can also be continued into the outer region $x>1$, but a different chart is required to reach the Cauchy horizon~\cite{MartinGarcia:2003,Ecker:2026}. The present work stays in the past patch, where (2) is regular and the spherical reduction is unambiguous.

Second, any periodic $\beta$ solves the leading-order large-$D$ system. Uniqueness of the echoing period is not a leading-order statement; it appears only after next-to-leading $1/D$ corrections are imposed and convexity conditions such as $f\le1$ are enforced~\cite{Ecker:2026}. Continuously self-similar members of the family, obtained by taking $\beta$ constant, are likewise allowed at leading order and differ from the Clark--Pimentel CSS solutions at finite $D$. A quantum correction to $\Delta$ therefore cannot be extracted from a first-order Polyakov source alone. It requires a joint expansion in $1/D$ and a suitably scaled quantum coupling, a point we return to in section~4.

A third remark concerns numerics at large $D$. Direct finite-difference constructions of critical solutions become increasingly stiff as $D$ grows, because radial gradients localize near the horizon on the scale $r_0/D$~\cite{Emparan:2013a,Ecker:2026}. The analytic family (3) turns that stiffness into an advantage: the same localization that defeats a uniform numerical grid is what permits a closed-form leading-order solution. The semiclassical problem inherits a related nonuniformity, now from the ratio $\widehat\kappa/\Phi$ rather than from $1/D$. Matched inner and outer regions are therefore the natural language for both expansions.

\section{Spherical reduction, Polyakov anomaly, and the regular quantum state}

\subsection{Classical reduced theory}

Reducing the $D$-dimensional Einstein--massless-Klein--Gordon action on the unit $S^n$ produces, in the direct radial--temporal frame,
\begin{equation}
S_{\mathrm{red}}=C_n\int\dd^2x\sqrt{-h}\left[\Phi R+U(\Phi)(\nabla\Phi)^2+V(\Phi)-\Phi(\nabla\psi)^2\right],
\end{equation}
with
\begin{equation}
C_n=\frac{\Omega_n}{16\pi G_D},\qquad \Phi=r^n,\qquad U(\Phi)=\frac{n-1}{n\Phi},\qquad V(\Phi)=n(n-1)\Phi^{1-2/n}.
\end{equation}
This is the standard spherical reduction of Einstein gravity~\cite{Grumiller:2002,Cavaglia:1999,Kummer:1997}, not a vacuum linear-dilaton model by itself. The outer background of section~2 is an exact classical solution of (14) in the large-$D$ limit. The warped-product curvature identities that produce (14) are recorded in appendix~A.

The invariant metric equation may be written as
\begin{equation}
C_nE_{ab}=T^P_{ab},
\end{equation}
where
\begin{align*}
E_{ab}={}&2(h_{ab}\Box_h-\nabla_a\nabla_b)\Phi+2U\,\partial_a\Phi\partial_b\Phi-h_{ab}U(\nabla\Phi)^2-h_{ab}V\\
&-2\Phi\,\partial_a\psi\partial_b\psi+h_{ab}\Phi(\nabla\psi)^2.
\end{align*}
The Klein--Gordon equation is $\nabla_a(\Phi\nabla^a\psi)=0$. In the absence of the Polyakov tensor these equations reproduce the large-$D$ system of Ref.~\cite{Ecker:2026}.

\subsection{Localization of the anomaly-induced action}

For $N$ conformal fields the nonlocal Polyakov action is~\cite{Polyakov:1981}
\begin{equation}
S_P=-\frac{N}{96\pi}\int\dd^2x\sqrt{-h}\,R\Box_h^{-1}R,\qquad \kappa:=\frac{N}{48\pi}.
\end{equation}
Introducing an auxiliary scalar $\chi$ through $\Box_h\chi=R$ yields the local action with the same sign and normalization,
\begin{equation}
S_{P,\mathrm{loc}}=-\kappa\int\dd^2x\sqrt{-h}\left[\frac12(\nabla\chi)^2+\chi R\right].
\end{equation}
Varying $\chi$ recovers $\Box_h\chi=R$, and eliminating $\chi$ recovers (17). Writing $T^P_{ab}=\kappa P_{ab}$, the on-shell stress tensor is
\begin{equation}
P_{ab}=\partial_a\chi\partial_b\chi-\frac12h_{ab}(\nabla\chi)^2+2h_{ab}R-2\nabla_a\nabla_b\chi.
\end{equation}
It satisfies the trace identities
\begin{equation}
P^a{}_a=2R,\qquad \langle T^a{}_a\rangle=2\kappa R=\frac{N}{24\pi}R.
\end{equation}
The reduced coupling that multiplies $P_{ab}$ in the metric equation (16) is
\begin{equation}
\widehat\kappa:=\frac{\kappa}{C_n}.
\end{equation}
All subsequent estimates are phrased in terms of $\widehat\kappa$ and the spherical dilaton $\Phi$.

Two remarks are in order. First, the overall sign in (18) is fixed by the requirement that the trace anomaly have the conventional positive coefficient for ordinary conformal matter. Reversing the sign would convert the Polyakov flux into an unphysical anti-evaporation source. Second, (18) is an intrinsic two-dimensional action. It does not automatically encode the dilaton-dependent terms generated by reducing higher-dimensional quantum matter on $S^n$~\cite{Frolov:2000,Balbinot:1999}. Those terms arise because the functional measure for a higher-dimensional field depends on the warp factor $r^n$, and they produce additional contributions to the trace anomaly of the schematic form $R f(\Phi)+(\nabla\Phi)^2g(\Phi)+V_q(\Phi)$. We treat them as a separate deformation and leave them for future study.

\subsection{Conservative equation and the zero-mode obstruction}

On the background (2) the auxiliary equation $\Box_h^{(0)}\chi_0=R_0$ takes the conservative form
\begin{align}
L_\beta\chi_0:={}&\partial_\tau\left[-\frac{\partial_\tau\chi_0+x\partial_x\chi_0}{f_0}\right]
+\partial_x\left[-\frac{x}{f_0}\partial_\tau\chi_0+\left(f_0-\frac{x^2}{f_0}\right)\partial_x\chi_0\right]
=S_\beta,\\
S_\beta(\tau,x)={}&-\frac{2\beta^2(\tau)}{\sqrt{B(\tau)}\,Q^{3/2}(\tau,x)}.
\end{align}
The left-hand side is a total divergence. This is what makes the global obstruction manifest.

Integrate (22) over one echo and over $0\le x\le1$. A strictly periodic auxiliary field would cancel the $\tau$-flux through the two ends of the fundamental domain. Smooth-center regularity, $\partial_x\chi_0(\tau,0)=0$, cancels the radial flux at $x=0$. At the SSH one has $f_0(\tau,1)=1$, and the remaining flux reduces to
\begin{equation}
J^x\big|_{x=1}=-\partial_\tau\chi_0(\tau,1).
\end{equation}
The integrated source, however, does not vanish. Strict periodicity would therefore require
\begin{equation}
0=\int_0^1\dd\tau\int_0^1\dd x\,S_\beta=-2\int_0^1\dd\tau\,\frac{\beta^2}{1+\beta^2},
\end{equation}
which is impossible for any nontrivial real $\beta$. This is a global Fredholm obstruction, not a numerical artifact and not an artifact of a particular gauge.

The consistent condition is additive monodromy,
\begin{equation}
\chi_0(\tau+1,x)=\chi_0(\tau,x)+C,\qquad C=2\int_0^1\dd\tau\,\frac{\beta^2}{1+\beta^2}.
\end{equation}
For the explicit profile (4) one finds
\begin{equation}
C=0.586305742069516.
\end{equation}
Writing $\chi_0=C\tau+\varphi$ with $\varphi(\tau+1,x)=\varphi(\tau,x)$ converts (22) into an equation for the periodic remainder,
\begin{equation}
L_\beta\varphi=S_\beta-C L_\beta\tau,\qquad L_\beta\tau=\frac{\partial_\tau f_0}{f_0^2}-\frac1{f_0}+\frac{x\partial_x f_0}{f_0^2}.
\end{equation}
The new source has vanishing integrated zero mode by construction. At $x=1$ the characteristic regularity condition becomes
\begin{equation}
B\varphi_{\tau\tau}+2B\varphi_{\tau x}+\varphi_\tau+2\varphi_x=\frac{2\beta^2}{B}-C.
\end{equation}
Although $\chi_0$ itself is only affine-periodic, its first derivatives and the tensor $P_{ab}$ remain periodic covariant objects. The stress tensor that sources the metric is therefore compatible with discrete self-similarity even though the auxiliary field is not.

\subsection{Null Floquet coordinates and the SSH-regular state}

The conservative analysis identifies the monodromy but does not yet select the quantum state. That selection is most transparent in conformal gauge. Using the radial null derivatives (9) on the leading background, define null coordinates $u,v$ by
\begin{equation}
D_+u=0,\qquad D_-v=0.
\end{equation}
DSS-adapted Floquet forms are
\begin{equation}
u=-\ee^{-\lambda\tau}p(\tau,x),\qquad v=-\ee^{-\lambda\tau}q(\tau,x),
\end{equation}
where $p$ and $q$ are periodic and normalized by $p(\tau,0)=q(\tau,0)=1$. Then
\begin{equation}
D_+p=\lambda p,\qquad D_-q=\lambda q,
\end{equation}
and one echo acts as $(u,v)\mapsto(\ee^{-\lambda}u,\ee^{-\lambda}v)$.

Near the SSH, with $y=1-x$,
\begin{equation}
x-f_0=-\frac{y}{B(\tau)}+O(y^2).
\end{equation}
If $q=a(\tau)y^\alpha+\cdots$, periodicity of $a$ implies
\begin{equation}
\alpha\int_0^1\frac{\dd\tau}{B(\tau)}=\lambda.
\end{equation}
A nondegenerate coordinate has $\alpha=1$, so SSH regularity fixes the Floquet exponent
\begin{equation}
\lambda=\int_0^1\frac{\dd\tau}{1+\beta^2(\tau)}=0.706847128965242
\end{equation}
for the profile (4). The SSH is then the regular null surface $v=0$. The same argument with the opposite null family determines the behavior of $p$ near $x=1$, but $u$ need not vanish on the SSH; only $v$ does. The exponent $\lambda$ is therefore a property of the outgoing (or, in the past patch, the inward-pointing) characteristic family, and it is fixed once and for all by the requirement that the SSH be a nondegenerate coordinate surface rather than an accumulation surface of log-periodic oscillations.

Write the background in conformal gauge as $\dd s_0^2=-\ee^{2\rho(u,v)}\dd u\dd v$. The transformation (31) gives
\begin{equation}
\ee^{2\rho}=-\frac{\ee^{-2\tau}}{u_xv_x}=\ee^{-2(1-\lambda)\tau}\frac1{-p_xq_x},
\end{equation}
and therefore
\begin{equation}
\rho=-(1-\lambda)\tau+\rho_{\mathrm{per}},\qquad \rho_{\mathrm{per}}:=-\frac12\ln(-p_xq_x),
\end{equation}
with $\rho_{\mathrm{per}}$ periodic. Note that the monodromy (26) and the Floquet exponent (35) imply the identity
\begin{equation}
C=2(1-\lambda).
\end{equation}
Consequently
\begin{equation}
\rho(\tau+1,x)=\rho(\tau,x)-\frac C2.
\end{equation}
The same affine shift is therefore obtained from a global integral identity and from the local scaling of a regular null chart. This is a nontrivial consistency check: the two calculations use different data, and there is no a priori reason for them to agree unless the SSH has been coordinatized in a regular way.

In conformal gauge one has
\begin{equation}
R=8\ee^{-2\rho}\rho_{uv},\qquad \Box_h=-4\ee^{-2\rho}\partial_u\partial_v.
\end{equation}
The general solution of $\Box_h\chi=R$ is
\begin{equation}
\chi=-2\rho+F(u)+G(v),
\end{equation}
where the chiral functions $F$ and $G$ specify the quantum state. Because $v=0$ is a nondegenerate SSH coordinate, regularity requires $G$ to be smooth there. Exact DSS covariance requires $G(\ee^{-\lambda}v)=G(v)$ up to a constant monodromy. A smooth solution of that functional equation is constant; every nonconstant solution is log-periodic and oscillates infinitely often as $v\to0$. At the regular center the even extension gives $p_x=-q_x$ and $\rho_x=0$. The condition $\chi_x=0$ then relates $F'$ and $G'$ on $u=v$. Simultaneous center and SSH regularity therefore fixes
\begin{equation}
\chi_{\mathrm{reg}}=-2\rho+\chi_*.
\end{equation}
This state automatically inherits the monodromy
\begin{equation}
\chi_{\mathrm{reg}}(\tau+1,x)=\chi_{\mathrm{reg}}(\tau,x)+C\rev{.}
\end{equation}
Substitution of (42) into (19) yields the regular-state Polyakov tensor
\begin{equation}
P_{uu}=4(\rho_{uu}-\rho_u^2),\qquad P_{vv}=4(\rho_{vv}-\rho_v^2),\qquad P_{uv}=-4\rho_{uv}.
\end{equation}
Nonconstant $F$ or $G$ would add state-dependent chiral fluxes to the first two components, corresponding to incoming or outgoing radiation that is singular at $v=0$ or incompatible with center regularity. The state (42) is the unique choice that avoids both defects.

\section{Semiclassical equations and the failure of a uniform outer expansion}

\subsection{Field equations in conformal gauge}

The full reduced action is $S=S_{\mathrm{red}}+S_{P,\mathrm{loc}}$. Using the SSH-regular state, the mixed metric equation becomes
\begin{equation}
2\Phi_{uv}+\frac12\ee^{2\rho}V(\Phi)+4\widehat\kappa\rho_{uv}=0.
\end{equation}
The two null constraints are
\begin{align}
-2\Phi_{uu}+4\rho_u\Phi_u+2U(\Phi)\Phi_u^2-2\Phi\psi_u^2&=4\widehat\kappa(\rho_{uu}-\rho_u^2),\nonumber\\
-2\Phi_{vv}+4\rho_v\Phi_v+2U(\Phi)\Phi_v^2-2\Phi\psi_v^2&=4\widehat\kappa(\rho_{vv}-\rho_v^2).
\end{align}
Variation with respect to $\Phi$ gives
\begin{equation}
2\rho_{uv}+2U(\Phi)\Phi_{uv}+U'(\Phi)\Phi_u\Phi_v+\psi_u\psi_v+\frac14\ee^{2\rho}V'(\Phi)=0,
\end{equation}
and the Klein--Gordon equation remains
\begin{equation}
2\Phi\psi_{uv}+\Phi_u\psi_v+\Phi_v\psi_u=0.
\end{equation}
Equations (45)--(48) constitute the closed semiclassical system. They are exact within the minimally coupled Polyakov model; no further large-$D$ reduction has been imposed at this stage.

\subsection{Critical kinetic determinant}

Equations (45) and (47) form a linear $2\times2$ system for the mixed derivatives $\Phi_{uv}$ and $\rho_{uv}$,
\begin{equation}
\begin{pmatrix}2&4\widehat\kappa\\2U&2\end{pmatrix}
\begin{pmatrix}\Phi_{uv}\\\rho_{uv}\end{pmatrix}
=\begin{pmatrix}A\\B_0\end{pmatrix},
\end{equation}
where
\begin{equation}
A=-\frac12\ee^{2\rho}V,\qquad B_0=-U'\Phi_u\Phi_v-\psi_u\psi_v-\frac14\ee^{2\rho}V'.
\end{equation}
The determinant of the coefficient matrix is $4\cD(\Phi)$, with
\begin{equation}
\cD(\Phi)=1-2\widehat\kappa U(\Phi)=1-\frac{\Phi_c}{\Phi},\qquad \Phi_c=2\widehat\kappa\frac{n-1}{n}.
\end{equation}
Away from $\Phi=\Phi_c$ one may solve
\begin{equation}
\Phi_{uv}=\frac{A-2\widehat\kappa B_0}{2\cD},\qquad \rho_{uv}=\frac{B_0-UA}{2\cD}.
\end{equation}
Define the compatibility defect
\begin{equation}
\cJ:=-U'\Phi_u\Phi_v-\psi_u\psi_v+\frac{\ee^{2\rho}}4\left(\frac{V}{\widehat\kappa}-V'\right).
\end{equation}
At $\Phi=\Phi_c$ a $C^2$-smooth crossing exists only if $\cJ|_{\Phi_c}=0$. Otherwise
\begin{equation}
\Phi_{uv}\sim-\frac{\widehat\kappa\cJ_c}{\cD},\qquad \rho_{uv}\sim\frac{\cJ_c}{2\cD},\qquad R\sim\frac{4\ee^{-2\rho_c}\cJ_c}{\cD}.
\end{equation}
The degeneracy is physical: it reflects a vanishing of the kinetic matrix of the reduced theory, not a coordinate singularity of the conformal chart. Chiral state functions modify the null constraints (46) but do not change the mixed compatibility condition $\cJ|_{\Phi_c}=0$. In that sense, the critical surface is independent of the chiral state dependence, which SSH regularity has already removed.

The same structure appears in other two-dimensional semiclassical models whenever the Polyakov coupling competes with the dilaton kinetic term~\cite{RST:1992,Vaz:1994}. What is special here is that $\Phi=r^n$ is tied to the areal radius of a higher-dimensional sphere. The critical value $\Phi_c$ therefore defines a critical radius $r_c:=\Phi_c^{1/n}$, and whether a DSS solution can cross that radius smoothly becomes a concrete geometric question about the center.

\subsection{Outer perturbation and the scale obstruction}

At fixed $\widehat\kappa$ a formal outer expansion reads
\begin{equation}
\rho=\rho_0+\widehat\kappa\rho_1+\cdots,\qquad \Phi=\Phi_0+\widehat\kappa\Phi_1+\cdots,\qquad \psi=\psi_0+\widehat\kappa\psi_1+\cdots.
\end{equation}
The $O(\widehat\kappa)$ mixed equation is, for example,
\begin{equation}
2(\Phi_1)_{uv}+\ee^{2\rho_0}\left[V(\Phi_0)\rho_1+\frac12V'(\Phi_0)\Phi_1\right]=-4(\rho_0)_{uv},
\end{equation}
and the $uu$ constraint is sourced by $4[(\rho_0)_{uu}-(\rho_0)_u^2]$, with an analogous $vv$ equation. Superficially this looks like a standard linear problem for periodic corrections on a periodic background.

The obstruction is the coupling, not the source. On the classical background
\begin{equation}
\Phi_0=r^n=\ee^{-n\tau}x^n,\qquad g_{\mathrm{eff}}(\tau,x):=\frac{\widehat\kappa}{\Phi_0}=\widehat\kappa\ee^{n\tau}x^{-n}.
\end{equation}
Under one echo $g_{\mathrm{eff}}$ grows by $\ee^n$. The Polyakov tensor is periodic as a covariant tensor, but its coupling to the spherical dilaton is not. A periodic first-order correction therefore cannot remain uniform at fixed quantum coupling.

The same obstruction is explicit in areal gauge. Recall
\begin{equation}
\dd s_2^2=-\alpha^2(t,r)\dd t^2+a^2(t,r)\dd r^2,\qquad f=\frac\alpha a,\qquad \Omega=\frac{(n-1)(a^2-1)}{x^2}.
\end{equation}
The radial equations~\cite{Ecker:2026} acquire sources from the Polyakov stress tensor,
\begin{equation}
x\partial_x\Omega+x^2\Omega^2=(\text{classical scalar terms})+S_\Omega^{(P)},
\end{equation}
with
\begin{equation}
S_\Omega^{(P)}=\frac{n-1}{n}g_{\mathrm{eff}}\frac{a^2}{f^2}\ee^{-2\tau}P_{tt}.
\end{equation}
The expansion is valid only where $g_{\mathrm{eff}}\ll1$ and inevitably fails as $r\to0$.

An ordinary Poincaré--Lindstedt shift of the echoing period also fails at this order. Its solvability condition presupposes a periodic source, whereas (57)--(60) are not periodic at fixed coupling. In addition, the period of the leading large-$D$ solution is a modulus: it is selected only by subleading $1/D$ consistency conditions~\cite{Ecker:2026}. A meaningful quantum shift of $\Delta$ therefore requires a joint expansion in $1/D$ and a suitably scaled quantum coupling. Treating $\widehat\kappa$ as an independent small parameter at infinite $D$ cannot determine $\Delta$.

The correct organization is matched asymptotics. Outside a quantum core of radius $r_c=\Phi_c^{1/n}$ one may still treat the Polyakov tensor perturbatively. Inside that core the coupling is of order unity and the equations must be kept nonlinear. The next section constructs that inner problem at leading order in the inner scale.

\section{Nonlinear inner problem, matching, and throat formation}

\subsection{Inner scale and frozen DSS phase}

A controlled overlap with the classical outer solution requires
\begin{equation}
r_c\ll r\ll-t,\qquad \text{or equivalently}\qquad
1\ll\frac r{r_c}\ll\frac1{x_c},\qquad x_c(\tau):=\frac{r_c}{-t}=\ee^\tau r_c.
\end{equation}
At the center the outer scalar has proper-time derivative
\begin{equation}
\omega(\tau):=\frac{\dd\psi}{\dd T}=\frac{\sqrt n\,\beta(\tau)}{-t}.
\end{equation}
The dimensionless inner matter strength is therefore
\begin{equation}
\mq(\tau):=\omega(\tau)r_c=\sqrt n\,\beta(\tau)x_c(\tau).
\end{equation}
When $x_c\ll1$, inner radial derivatives scale as $r_c^{-1}$ while phase evolution scales as $(-t)^{-1}=x_cr_c^{-1}$. Freezing the DSS phase is the leading inner approximation. The weak-matter expansion below further assumes $|\mq|\ll1$, which is automatic at sufficiently early echoes for any fixed $\widehat\kappa$ of semiclassical size.

The frozen-phase reduction is not uniformly valid near the isolated zeros of $\beta$. There $\mq$ vanishes and time derivatives, elsewhere suppressed by $x_c$, become essential inside narrow phase layers. We return to those layers at the end of this section and in section~6. Away from the zeros the frozen-phase system already exhibits the mechanism that prevents a regular center.

\subsection{Frozen-phase equations}

Use local static conformal coordinates at fixed phase,
\begin{equation}
\dd s_2^2=\ee^{2\rho(z)}(-\dd T^2+\dd z^2),\qquad \Phi=\Phi(z),\qquad \psi=\omega T.
\end{equation}
The Klein--Gordon equation is then automatic. The remaining equations reduce to
\begin{align}
\Phi''+2\widehat\kappa\rho''&=\ee^{2\rho}V,\\
-2\rho''-2U\Phi''-U'(\Phi')^2+\omega^2+\ee^{2\rho}V'&=0,\\
2\widehat\kappa(\rho')^2+2\rho'\Phi'+U(\Phi')^2-\Phi\omega^2-\ee^{2\rho}V&=0.
\end{align}
The two null constraints coincide in this sector. Defining $Y:=\Phi'+2\widehat\kappa\rho'$ produces the first-order system
\begin{align}
Y'&=\ee^{2\rho}V,\\
\cD(\Phi)(\Phi')^2&=Y^2-2\widehat\kappa(\ee^{2\rho}V+\Phi\omega^2),\\
\rho'&=\frac{Y-\Phi'}{2\widehat\kappa}.
\end{align}
Equations (68)--(70), with a continuous choice of square-root branch in (69), define the nonlinear inner solution by quadrature.

A smooth crossing of $\Phi_c$ must satisfy both
\begin{equation}
(\Phi_c')^2=2\widehat\kappa(\ee^{2\rho_c}V_c-\Phi_c\omega^2),\qquad
Y_c^2=2\widehat\kappa(\ee^{2\rho_c}V_c+\Phi_c\omega^2).
\end{equation}
These are extra shooting conditions. They hold for the exact vacuum branch $\omega=0$, $\rho=0$, $\Phi=z^n$, but not for generic time-dependent matter data. The vacuum solution crosses $\Phi_c$ because the compatibility defect $\cJ$ vanishes identically. Any nonzero $\omega$ detunes that cancellation.

\subsection{Regular-center, zero-mass expansion}

Normalize $\Phi_c=r_c^n$ and set $s:=z/r_c$. Choose the normalization of $z$ so that $\dd z/\dd r\to1$ in the overlap; hence $s\sim r/r_c$ there, although the two ratios are not identical away from the overlap. For $|\mq|\ll1$, the branch with no Schwarzschild-mass or singular-center homogeneous mode has
\begin{equation}
\frac{\Phi}{\Phi_c}=s^n\left[1+\mq^2\Ain(s)+O(\mq^4)\right],\qquad
\rho=\mq^2\Bin(s)+O(\mq^4).
\end{equation}
Classical overlap fixes
\begin{equation}
\Ain(s)\sim\frac{n-1}{6(n+1)}s^2,\qquad
\Bin(s)\sim\frac1{2(n+1)}s^2,\qquad s\to\infty,\quad |\mq|s\ll1.
\end{equation}
For $n\ge4$ the complete zero-mass asymptotic series can be written
\begin{equation}
\Ain=\sum_{j\ge0}a_js^{m_j},\qquad \Bin=\sum_{j\ge0}b_js^{m_j},\qquad m_j=2-jn,
\end{equation}
with
\begin{equation}
a_0=\frac{n-1}{6(n+1)},\qquad b_0=\frac1{2(n+1)}.
\end{equation}
The coefficients obey
\begin{equation}
(n-1)(m_j+1)a_j+n[m_j-(n-1)]b_j=\frac12\delta_{j0},
\end{equation}
and, for $j\ge1$,
\begin{equation}
[m_j(m_j-1)+2nm_j+2(n-1)]a_j-2n(n-1)b_j
=-\frac n{n-1}m_{j-1}(m_{j-1}-1)b_{j-1}.
\end{equation}
The homogeneous powers $s^0$, $s^{-1}$, and $s^{1-n}$ are set to zero by the regular-center, zero-mass match. The exceptional case $n=3$ contains a resonant logarithm and lies outside the large-$D$ regime considered here.

Analytic continuation of the same series toward the critical surface yields
\begin{equation}
\Ain(s)=-\frac1{n^2-1}\ln(s-1)+O(1),\qquad
\Bin(s)=\frac1{n(n+1)}\ln(s-1)+O(1).
\end{equation}
The weak-matter expansion is therefore nonuniform at $s=1$. These logarithms are the \rev{linearized signal} of a narrow nonlinear layer, derived independently by a boundary-layer analysis in appendix~B.

\subsection{Throat formation}

Numerical integration of (65)--(67) inward from the asymptotic series produces a turning point $\Phi'=0$ before $\cD$ vanishes. Representative dimensionless results are collected in table~1. The small-$\mq$ limit is
\begin{equation}
\frac{\Phi_{\min}}{\Phi_c}=1+\frac{2\mq^2}{n^2-1}+o(\mq^2).
\end{equation}
At the turning point,
\begin{equation}
R_{\mathrm{th}}r_c^2\sim\frac{n(n-1)(n^2-1)}{2\mq^2},\qquad \mq\to0.
\end{equation}
For every \rev{finite, nonzero $\mq$} the ODE coefficients remain finite because $\Phi_{\min}>\Phi_c$. Nevertheless the solution no longer has a regular spherical center: the areal dilaton has a positive minimum and \rev{continues on to} a second radial branch. The matched endpoint is a local throat, or cornucopion, of the type familiar in semiclassical dilaton gravity~\cite{Banks:1992,Lowe:1993,Thorlacius:1995}. In those models a quantum-supported minimum of the dilaton replaces the would-be classical singularity of an evaporating two-dimensional black hole. The geometry continues through the throat onto a second asymptotic region, or else ends in a high-curvature tip, depending on the global boundary conditions. The present construction is local rather than global: we only match through the overlap (61) and do not specify the second-branch asymptotics. What matters for DSS collapse is that the first branch, the one that can be joined to the classical exterior, never reaches a regular center.

The curvature (80) diverges as $\mq\to0$. In that limit the throat is pressed against the critical surface and the boundary layer collapses. For any fixed nonzero $\mq$ the curvature is large but finite, of order $\mq^{-2}$ in units of $r_c^{-2}$. The inner solution therefore resolves the classical center into a high-curvature quantum core whose proper size is a small fraction of $r_c$.

\begin{table}[t]
\centering
\caption{Representative integrations of the frozen-phase inner system. The last column approaches $2/(n^2-1)$ as $\mq\to0$, in agreement with (79).}
\begin{tabular}{cccc}
\toprule
$n$ & $\mq$ & \rev{$s_{\mathrm{th}}=z_{\mathrm{th}}/r_c$} & $(\Phi_{\min}/\Phi_c-1)/\mq^2$\\
\midrule
4&0.030&0.99967952&0.133461\\
4&0.003&0.99999611&0.133335\\
10&0.030&0.99997380&0.021993\\
10&0.003&0.99999971&0.020220\\
\bottomrule
\end{tabular}
\end{table}

\subsection{Branches and loss of uniformity}

The inner problem admits three distinct continuations.
\begin{enumerate}
\item Vacuum fine-tuning. For $\omega=0$ the exact solution $\rho=0$, $\Phi=z^n$ crosses $\Phi_c$ because $\cJ=0$ identically.
\item Generic crossing. If a solution reaches $\Phi_c$ with $\cJ\ne0$, equation (54) produces a curvature singularity.
\item Regular outer matching. The regular-center, zero-mass outer coefficients select the throat (79), avoiding the determinant zero but losing the regular center.
\end{enumerate}

The explicit function $\beta(\tau)$ has isolated zeros. Since $\mq\propto\beta$, the frozen-phase throat becomes nonuniform near those phases, as is already visible from (80). Time derivatives, which are suppressed away from the zeros, become essential in narrow phase layers around them. They may influence the local choice between the throat and singular branches, but a smooth crossing still requires the extra condition (53) along the entire critical curve. Isolated zeros of $\beta$ do not remove that obstruction; they only invalidate the frozen-phase reduction in \rev{parametrically narrow phase intervals}.

The equations also admit a formal constant-dilaton vacuum below the critical surface,
\begin{equation}
\Phi_*=\widehat\kappa\left(1-\frac2n\right),\qquad R_*=-\frac{V(\Phi_*)}{\widehat\kappa},\qquad \frac{\Phi_*}{\Phi_c}=\frac{n-2}{2(n-1)}<1.
\end{equation}
This is an $\mathrm{AdS}_2$-core candidate. A generic DSS branch cannot reach it \rev{because} the critical surface lies in between and the compatibility defect does not vanish on that surface.

\section{Semiclassical null-energy structure}

\subsection{Effective radial NEC}

The reduced scalar stress tensor has null components
\begin{equation}
T_{uu}^{(\psi)}=2C_n\Phi\psi_u^2,\qquad T_{vv}^{(\psi)}=2C_n\Phi\psi_v^2.
\end{equation}
Adding the regular-state Polyakov tensor (44) and dividing by the positive factor $2C_n$ defines
\begin{equation}
N_u:=\Phi\psi_u^2+2\widehat\kappa(\rho_{uu}-\rho_u^2),\qquad
N_v:=\Phi\psi_v^2+2\widehat\kappa(\rho_{vv}-\rho_v^2).
\end{equation}
\rev{The $u$- and $v$-directed radial NECs are saturated on $N_u=0$ and $N_v=0$, respectively.} These curves must be distinguished from the classical matter-flux curves (11). The anomaly contributes derivatives of the conformal factor that have no classical counterpart, so the two families need not coincide even perturbatively.

\subsection{Local splitting of a classical line}

Let $y$ be a normal coordinate to a classical $u$-directed flux line and suppose
\begin{equation}
\psi_u=a_u y+O(y^2).
\end{equation}
Evaluating the \rev{reduced Polyakov flux combination $\Pu:=\rho_{uu}-\rho_u^2$} on the classical line gives
\begin{equation}
N_u=\Phi a_u^2y^2+2\widehat\kappa\Pu+O(y^3,\widehat\kappa y,\widehat\kappa^2).
\end{equation}
If $\Pu<0$, the double zero splits into two roots
\begin{equation}
y_\pm=\pm\sqrt{\frac{-2\widehat\kappa\Pu}{\Phi a_u^2}}+O\!\left(\frac{\widehat\kappa}{\Phi}\right),
\end{equation}
and the strip between them violates the $u$-directed NEC. If $\Pu>0$, there is no nearby real root and the classical saturation line disappears. Higher-order terms decide the degenerate case $\Pu=0$.

The characteristic displacement scales as
\begin{equation}
\delta y_P\sim\sqrt{\frac{\widehat\kappa}{\Phi}}=O(\sqrt{g_{\mathrm{eff}}}).
\end{equation}
The Polyakov-induced displacement of an NEC line scales as $\sqrt{g_{\mathrm{eff}}}$. By comparison, the separation between the two classical NEC branches first appears at $O(\epsilon)$, whereas their bending first appears at $O(\epsilon^2)$. Therefore, the condition $\sqrt{g_{\mathrm{eff}}}\sim\epsilon$ marks the scale at which the quantum splitting competes with the classical branch separation, while $\sqrt{g_{\mathrm{eff}}}\sim\epsilon^2$ marks the scale at which it competes with the classical bending. In particular, the Polyakov correction can dominate the NNLO classical bending even when $g_{\mathrm{eff}}\ll1$, where the outer semiclassical expansion remains perturbatively controlled. Thus, well before the nonlinear inner region is reached, quantum backreaction may already split, eliminate, or broaden the classical NEC-saturation lines.

\subsection{Loss of the curvature coincidence}

Three curve families now have independent definitions, summarized in table~2. Their coincidence in (12) is a classical property. The Polyakov null flux contains independent derivatives of the conformal factor, and the trace anomaly (20) modifies the trace relation. Moreover, an intrinsic two-dimensional Polyakov action does not determine the angular pressures required for a unique $D$-dimensional uplift. Equation (83) is therefore the unambiguous reduced radial statement.

\begin{table}[h]
\centering
\caption{Distinct curve families in the semiclassical problem. Their coincidence is not preserved by the anomaly-induced stress tensor.}
\begin{tabular}{p{0.37\linewidth}p{0.53\linewidth}}
\toprule
Curve family&Defining equation\\
\midrule
Classical matter flux&\rev{$\Pi-x\Psi=0\leftrightarrow\psi_u=0$; $\Pi+x\Psi=0\leftrightarrow\psi_v=0$}\\
Effective radial NEC&$N_u=0$ or $N_v=0$\\
Geometric zero curvature&\rev{$R[h]=0$}, or $R_D=0$ after an uplift prescription\\
\bottomrule
\end{tabular}
\end{table}

\subsection{NEC violation near the throat}

For the static inner solution, choose $u=T-z$ and $v=T+z$. Then
\begin{equation}
\psi_u=\psi_v=\frac\omega2,\qquad 4(\rho_{uu}-\rho_u^2)=4(\rho_{vv}-\rho_v^2)=\rho''-(\rho')^2.
\end{equation}
Hence
\begin{equation}
N_u=N_v=\frac{\Phi\omega^2}{4}+\frac{\widehat\kappa}{2}[\rho''-(\rho')^2].
\end{equation}
The boundary-layer result (80) implies
\begin{equation}
\rho''_{\mathrm{th}}r_c^2\sim-\frac{n(n-1)(n^2-1)}{4\mq^2}.
\end{equation}
The scalar term in (89) scales as $\mq^2$, whereas the negative Polyakov contribution scales as $-\mq^{-2}$. Both radial NECs are therefore violated near the throat when $|\mq|\ll1$.

Moving outward, the anomaly term weakens and the positive scalar flux eventually dominates. Using the weak-inner expansion (72), the leading outer edge of the NEC-violating layer obeys
\begin{equation}
\frac{n-1}{n}s^n+\Bin''(s)=0.
\end{equation}
Using the critical logarithm (78) gives
\begin{equation}
s_{\mathrm{NEC}}-1\simeq\frac1{\sqrt{n^2-1}}.
\end{equation}
Thus the effective NEC curves bound a negative-energy layer surrounding the quantum-supported throat. The derivation is given in appendix~B.

\subsection{Vertices, phase layers, and numerical diagnostics}

At a classical vertex $\beta(\tau_0)=0$, the frozen-phase parameter $\mq(\tau_0)$ vanishes and the throat expansion becomes nonuniform. Meanwhile, phase derivatives in $\rho_{uu}$ and $\rho_{vv}$ can remain finite through terms proportional to $[\beta'(\tau_0)]^2$. Locally, the time-dependent inner solution may split each classical null-flux branch into a pair of effective NEC boundaries when its on-line Polyakov flux is \rev{negative, as illustrated in Fig.~1}, while the classical zero-curvature contours remain on the unsplit matter-flux lines. We remark that if the corresponding Polyakov flux is positive, that entire pair has no nearby real roots and the classical branch disappears from the effective NEC diagram.

\begin{figure}[t]
\centering
\includegraphics[]{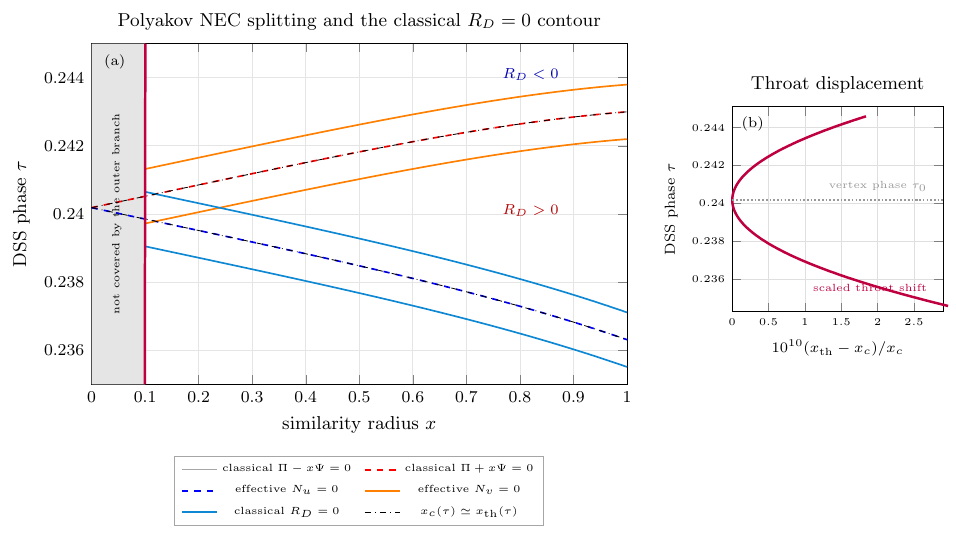}
\caption{\rev{Parameterized Polyakov splitting of the radial NEC-saturation curves for $D=300$ ($n=298$ and $\epsilon=1/299$). Panel (a): the red upper and blue lower dashed curves are the classical matter-flux lines $\Pi-x\Psi=0$ and $\Pi+x\Psi=0$, respectively. They coincide with the classical zero-curvature contour $R_D=0$, shown by the black dash-dotted curves. Assuming negative on-line Polyakov flux, $\Pu,\Pv<0$, the upper classical double zero splits into two $N_u=0$ boundaries and the lower classical double zero splits into two $N_v=0$ boundaries. The gray region $x<x_{\rm th}(\tau)$ is not covered by the regular-center outer branch; the nearly coincident critical-dilaton and throat curves are shown in purple. Panel (b): the single purple curve is the scaled positive fractional areal displacement $10^{10}(x_{\rm th}-x_c)/x_c=10^{10}[(\Phi_{\min}/\Phi_c)^{1/n}-1]$; the gray dotted horizontal line only marks the vertex phase $\tau=\tau_0$. We set $x_c(\tau_0)=0.1$ and use the explicit schematic prescription $\tau_{u,\pm}=\tau_+^{\rm cl}\pm\sigma$ and $\tau_{v,\pm}=\tau_-^{\rm cl}\pm\sigma$, with $\sigma=8.0\times10^{-4}$.}}
\end{figure}

\section{Discussion and conclusions}

The analysis above uses a minimally coupled intrinsic two-dimensional Polyakov action in the direct spherical-reduction frame. When that stress tensor is formally uplifted, it corresponds to a $D$-dimensional energy density proportional to $P_{ab}/r^n$ and is singular at a regular spherical center. This is not a defect of the calculation so much as a statement about the model: the intrinsic two-dimensional conformal theory and the dimensionally reduced higher-dimensional quantum theory are different semiclassical models. The latter generally produces additional dilaton-dependent anomaly terms~\cite{Frolov:2000,Balbinot:1999,Nojiri:1999}. Those terms can change both the location of the critical surface and the structure of the inner kinetic matrix. They are the natural next deformation.

Within the model studied here the conclusions are robust and can be stated cleanly.
\begin{itemize}
\item The auxiliary field is affine-periodic, with monodromy $C=2(1-\lambda)$. The same value of $C$ follows from a global zero-mode integral and from the affine scaling of a regular conformal factor.
\item Simultaneous center and SSH regularity uniquely selects $\chi=-2\rho+\chi_*$. Nonconstant chiral state functions are log-periodic at the SSH and are incompatible with a regular null chart.
\item The outer quantum expansion fails when $\widehat\kappa/\Phi=O(1)$. The obstruction is the growing ratio of the Polyakov coupling to the spherical dilaton, not a failure of discrete self-similarity of the stress tensor itself.
\item Generic crossings of $\Phi_c$ are singular. The compatibility defect $\cJ$ does not vanish for generic matter data.
\item Regular outer matching selects a high-curvature throat rather than a regular center. The throat is surrounded by a layer of radial NEC violation whose outer edge sits a coordinate distance of order $(n^2-1)^{-1/2}$ beyond the critical radius.
\item A numerical shift of the echoing period cannot be obtained before the subleading $1/D$ period-selection operator and the quantum inner problem are treated together.
\end{itemize}

Recent semiclassical studies of quantum critical collapse by Toma\v{s}evi\'c and Wu~\cite{TomasevicWu:2026PRD,TomasevicWu:2026JHEP} find a quantum growing mode, a finite mass gap, and quantum-generated trapped surfaces in near-critical Einstein--scalar collapse. These results do not contradict the present high-curvature throat: those works analyze horizon formation and global visibility in specific $2+1$- and $3+1$-dimensional settings, whereas we study the local inner geometry of large-$D$ DSS solutions with an intrinsic two-dimensional Polyakov action. Our throat construction neither establishes nor excludes a horizon in a more complete semiclassical evolution.

These results, however, do not yet determine whether a more complete quantum theory censors, evaporates, or preserves the classical singularity. Two further directions are visible already. First, the CSS members of the Ecker--Ecker--Grumiller family, corresponding to constant $\beta$, have no phase layers. They provide a technically simpler inner problem in which the throat mechanism can be isolated without the additional complication of echoing zeros. Second, the large-$D$ expansion of Hawking radiation and of the tunneling rate~\cite{Emparan:2013a,Wei:2014} suggests that the quantum flux through the SSH could be organized as a $1/D$ series of its own. Combining that flux with the inner throat constructed here \rev{would test quantitatively whether, and how, the critical solution evaporates}.

From the point of view of critical collapse, the throat is a qualitatively new object. Classically, the past-patch center is regular at every finite $\tau$ and only becomes singular as $\tau\to+\infty$. Semiclassically, already at finite $\tau$, the areal radius cannot fall below $r_c$ on the matched branch. The price is a curvature of order $\mq^{-2}r_c^{-2}$ and a layer of NEC violation. Whether that core subsequently evaporates, bounces, or settles into an $\mathrm{AdS}_2$ remnant is not decided by the frozen-phase equations. It is decided by the time-dependent inner problem near the zeros of $\beta$ and by \rev{the additional dilaton-dependent anomaly terms required by the underlying higher-dimensional quantum theory}.

The minimally coupled Polyakov model does not yield a generic smooth, exactly self-similar completion of the classical center. That negative statement is useful. It shows that any such completion, if it exists, must come from physics that is not contained in the intrinsic two-dimensional conformal anomaly alone.

\section*{Acknowledgements}
WYW is supported in part by the National Science and Technology Council of Taiwan under Grants No.~114-2112-M-033-011 and 115-2112-M-033-010. SHL is supported in part by Grant No.~114-2112-M-033-014-MY3.

\section*{Declaration of generative AI use}
A language model was used to assist with the presentation and preliminary checking of selected mathematical derivations. All equations, intermediate steps, and conclusions were independently derived or verified by the authors. The authors take full responsibility for the content.

\appendix
\section{Checks of the reduced and anomaly-induced actions}

\subsection{Warped-product reduction}

For the warped product
\begin{equation}
\dd s_D^2=h_{ab}(x)\dd x^a\dd x^b+r^2(x)\dd\Omega_n^2,
\end{equation}
the volume element and scalar curvature are
\begin{equation}
\sqrt{-g_D}=\sqrt{-h}\,r^n\sqrt{\gamma_n},
\end{equation}
\begin{equation}
R_D=R[h]+\frac{n(n-1)}{r^2}-\frac{2n}{r}\Box_h r-\frac{n(n-1)}{r^2}(\nabla r)^2.
\end{equation}
After angular integration, the term containing $\Box_hr$ may be integrated by parts,
\begin{equation}
\int\sqrt{-h}\,(-2nr^{n-1}\Box_hr)=\int\sqrt{-h}\,2n(n-1)r^{n-2}(\nabla r)^2,
\end{equation}
up to a boundary term. Setting $\Phi=r^n$ then gives
\begin{equation}
n(n-1)r^{n-2}(\nabla r)^2=\frac{n-1}{n\Phi}(\nabla\Phi)^2,
\qquad n(n-1)r^{n-2}=n(n-1)\Phi^{1-2/n},
\end{equation}
which proves (14) and (15).

\subsection{Polyakov localization}

Varying (18) with respect to $\chi$ gives
\begin{equation}
\delta_\chi S_{P,\mathrm{loc}}=-\kappa\int\sqrt{-h}\,(-\Box_h\chi+R)\delta\chi,
\end{equation}
and hence $\Box_h\chi=R$. On shell,
\begin{equation}
\int\sqrt{-h}\,(\nabla\chi)^2=-\int\sqrt{-h}\,\chi\Box_h\chi=-\int\sqrt{-h}\,\chi R.
\end{equation}
It follows that
\begin{equation}
S_{P,\mathrm{loc}}\big|_{\mathrm{on\ shell}}=-\frac\kappa2\int\sqrt{-h}\,\chi R
=-\frac{N}{96\pi}\int\sqrt{-h}\,R\Box_h^{-1}R.
\end{equation}
The metric variation uses the two-dimensional identity $R_{ab}-\tfrac12h_{ab}R=0$ and gives (19). Taking its trace and imposing $\Box_h\chi=R$ yields (20).

\section{Boundary-layer derivation of the throat and the NEC edge}

\subsection{Throat}

\rev{Measure lengths in units of $r_c$, so $r_c=1$ and hence $\Phi_c=r_c^n=1$.} The critical-dilaton relation then fixes
\begin{equation}
\widehat\kappa=\frac{n}{2(n-1)}.
\end{equation}
Resolve the nonuniform region signaled by (78) with
\begin{equation}
s=1+\mq^2\xi,\qquad \Phi=1+\mq^2F(\xi),\qquad \rho=\mq^2G(\xi).
\end{equation}
Since $\dd/\dd s=\mq^{-2}\dd/\dd\xi$, the leading part of (65) is
\begin{equation}
F_{\xi\xi}+2\widehat\kappa G_{\xi\xi}=0.
\end{equation}
Matching to the vacuum slope $\Phi'=n$ gives
\begin{equation}
F_\xi+2\widehat\kappa G_\xi=n.
\end{equation}
Writing $Y=n+\mq^2y(\xi)$, equation (68) yields $y_\xi=V_c=n(n-1)$. Expansion of (69) through order $\mq^2$ gives
\begin{equation}
F(F_\xi)^2=n^2F+K=n^2(F-F_{\mathrm{th}}),
\end{equation}
where $K=-n^2F_{\mathrm{th}}$. Thus
\begin{equation}
(F_\xi)^2=n^2\left(1-\frac{F_{\mathrm{th}}}{F}\right).
\end{equation}
For large positive $\xi$,
\begin{equation}
F=n\xi-\frac{F_{\mathrm{th}}}{2}\ln\xi+O(1).
\end{equation}
Matching the logarithm to (78) fixes
\begin{equation}
F_{\mathrm{th}}=\frac{2}{n^2-1},
\end{equation}
which gives (79). Differentiating (106) and evaluating at the turning point gives
\begin{equation}
F_{\xi\xi}^{\mathrm{th}}=\frac{n^2(n^2-1)}4,\qquad
G_{\xi\xi}^{\mathrm{th}}=-\frac{n(n-1)(n^2-1)}4.
\end{equation}
Since $R=-2\ee^{-2\rho}\rho''$, restoring $r_c$ produces (80).

\subsection{Outer NEC edge}

Use
\begin{equation}
\Phi=\Phi_cs^n+O(\mq^2),\qquad \omega=\frac{\mq}{r_c},\qquad
\rho=\mq^2\Bin(s)+O(\mq^4),\qquad \frac\dd{\dd z}=\frac1{r_c}\frac\dd{\dd s}.
\end{equation}
Neglecting $(\rho')^2=O(\mq^4)$, equation (89) becomes
\begin{equation}
N_u=\frac{\Phi_c\mq^2}{4r_c^2}\left[s^n+\frac{2\widehat\kappa}{\Phi_c}\Bin''(s)\right]+O(\mq^4).
\end{equation}
Using $\Phi_c=2\widehat\kappa(n-1)/n$ gives
\begin{equation}
\frac{n-1}{n}s^n+\Bin''(s)=0,
\end{equation}
which proves (91). Near $s=1$,
\begin{equation}
\Bin''(s)=-\frac1{n(n+1)(s-1)^2}+\cdots.
\end{equation}
Approximating $s^n\simeq1$ then gives $(s_{\mathrm{NEC}}-1)^2=1/(n^2-1)$ and (92).

\end{document}